# Thermoluminescence measurements of neutron streaming through JET Torus Hall ducts

Barbara Obryk[a*], Paola Batistoni[b,d], Sean Conroy[c,d], Brian D. Syme[d], Sergey Popovichev[d], Ion E. Stamatelatos[e],

Theodora Vasilopoulou[e], Paweł Bilski[a] and JET EFDA Contributors[*]

*EFDA JET, Culham Science Centre, Abingdon, Oxfordshire, OX14 3DB United Kingdom*

[a]*Institute of Nuclear Physics Polish Academy of Sciences, Radzikowskiego 152, 31-342 Kraków, Poland*

[b]*Associazione EURATOM-ENEA sulla Fusione, Via Enrico Fermi 45, 00044 Frascati, Rome, Italy*

[c]*EURATOM-VR Association, Dept. of Physics and Astronomy, Uppsala University, Box 516, 75120 Uppsala, Sweden*

[d]*EURATOM–CCFE Fusion Association, Culham Science Centre, Abingdon OX14 3DB, UK*

[e]*Institute of Nuclear and Radiological Sciences, Energy, Technology and Safety, NCSR "Demokritos", Athens, Greece*

[*]*See the Appendix of F. Romanelli et al., Proceedings of the 24th IAEA Fusion Energy Conference 2012, San Diego, USA*

**HIGHLIGHTS:**

• Thermoluminescence detectors (TLD) were used for dose measurements at JET.
• Pairs of 6LiF/7LiF TLDs allow to measure thermal neutron component of a radiation field.
• For detection of neutrons of higher energy, polyethylene (PE-300) moderators were used.
• TLDs were installed at eleven positions in the JET hall and the hall labyrinth.
• The experimental results are compared with calculations using the MCNP code.

**Abstract**

Thermoluminescence detectors (TLD) were used for dose measurements at JET. Several hundreds of LiF detectors of various types, standard LiF:Mg,Ti and highly sensitive LiF:Mg,Cu,P were produced. LiF detectors consisting of natural lithium are sensitive to slow neutrons, their response to neutrons being enhanced by $^6$Li-enriched lithium or suppressed by using lithium consisting entirely of $^7$Li. Pairs of $^6$LiF/$^7$LiF detectors allow distinguishing between neutron/non-neutron components of a radiation field. For detection of neutrons of higher energy, polyethylene (PE-300) moderators were used. TLDs, located in the centre of cylindrical moderators, were installed at eleven positions in the JET hall and the hall labyrinth in July 2012, and exposure took place during the last two weeks of the experimental campaign. Measurements of the gamma dose were obtained for all positions over a range of about five orders of magnitude variation. As the TLDs were also calibrated in a thermal neutron field, the neutron fluence at the experimental position could be derived. The experimental results are compared with calculations using the MCNP code. The results confirm that the TLD technology can be usefully applied to measurements of neutron streaming through JET Torus Hall ducts.

* Corresponding author: e-mail: barbara.obryk@ifj.edu.pl; tel.: +48-12-6628280; fax: +48-12-6628066

# 1. Introduction

Neutron streaming through penetrations of ITER structural and shielding materials is important for the safety assessment of the ITER biological shield. In particular, evaluation of neutron streaming outside the ITER biological shield through large ducts is a major safety task involving computations using state-of-the-art codes of radiation transport along long paths and in complex geometries. Therefore, a study performed at JET aiming at validating the calculation of neutron streaming through ducts and of the dose rates outside of the JET torus hall would be of outmost importance, since it would enable validation of the safety assessment calculations made for ITER.

For this purpose, thermoluminescence detectors (TLD) were used for dose measurements at JET. TLDs are well developed technology in the field of passive radiation sensors. Among them, very popular are lithium fluoride TL detectors [1]. The MCP (LiF:Mg,Cu,P) detectors, due to their very high sensitivity and a simple signal to dose relation, are now becoming standard in modern environmental thermoluminescence (TL) dosimetry [2]. They are able to measure absorbed doses at microgray levels and even below [3]. Based on the newly-discovered behaviour of LiF:Mg,Cu,P detectors at high and ultra-high doses [4,5], a new method of TL measurement of radiation doses ranging from micrograys up to a megagray, has been recently developed at the Institute of Nuclear Physics (IFJ) in Krakow, Poland. The method is based on the relationship between the TL signal, integrated in the given temperature range, and absorbed dose [6]. The 'ultra-high-temperature ratio' UHTR was defined in order to quantify the observed changes of the LiF:Mg,Cu,P glow-curve shape at very high doses and very high temperatures, which enabled determining an absorbed dose in the range from 1 kGy to 1 MGy. Thanks to this, the MCP (LiF:Mg,Cu,P) detector can measure doses ranging from below 1 µGy to about 1 MGy, also in mixed radiation fields. This newly established dosimetric method was tested in a range of radiation qualities, such as gamma radiation, electron and proton beams, thermal neutron fields and in high-energy mixed fields around the SPS and PS accelerators at CERN [7-9]. The TLD measurements can be used as a benchmark for the numerical evaluations [10].

The scope of the present work was the measurement and calculation of absorbed dose and of neutron fluence at the JET Torus Hall and its ducts during operation. Measurements have been performed using LiF TL detectors developed and produced at the IFJ in Kraków. The results of the measurements were compared against calculations using MCNP code conducted by the Institute of Nuclear and Radiological Sciences, Athens, Greece and by JET Neutron Group.

## 2. Materials and measurements

### 2.1 Materials

Several hundred lithium fluoride detectors of various types, made by the sintering technique, were prepared for measurements. Efforts concentrated on highly sensitive LiF:Mg,Cu,P (MCP-N), $^7$LiF:Mg,Cu,P (MCP-7), $^6$LiF:Mg,Cu,P (MCP-6) detectors, however, standard LiF:Mg,Ti (MTS-N), $^7$LiF:Mg,Ti (MTS-7), $^6$LiF:Mg,Ti (MTS-6) were also tested in order to observe differences in their response due to their higher efficiency to high-LET radiation. $^6$Li abundance in natural lithium is 7.59%, $^6$Li-enriched lithium contains 95.58% of $^6$Li while $^6$Li-suppressed lithium only 0.03% of $^6$Li. All MCP-N, MCP-7, MTS-N, MTS-7 detectors used were of typical size: 4.5 mm diameter and 0.9 mm thickness, while MCP-6 and MTS-6 detectors used were 4.5 mm diameter but 0.6 mm thickness only. All were developed and produced at the Radiation Physics and Dosimetry Department of the IFJ.

The most apparent difference between both detector types is obviously in their sensitivity to radiation, which is approximately 30 times higher for LiF:Mg,Cu,P than for LiF:Mg,Ti for gamma radiation [1]. The detection threshold of MCP detectors is below 1 μGy while for MTS it is only in the range of 20-50 μGy. The linearity range for both materials is at the level of a few Gy, while saturation dose is about 1 kGy. However, due to the newly discovered MCPs' high-dose high-temperature emission they are able to measure doses up to 1 MGy [5,6]. Another important difference between the dosimetric properties of these phosphors is in their dose response. MTS features the well-known linear-supralinear response, while MCP is linear-sublinear. The sublinear dose response of MCP bears some further consequences. It is generally accepted that this feature is responsible for the much lower TL efficiency with which heavy charged particles and high-LET particles are detected by MCP [11].

Due to differences in the neutron capture cross section of $^6$Li and $^7$Li isotopes, it is possible to detect thermal and epithermal neutrons with LiF detectors. $^6$Li has a high cross-section for low energy neutrons (about 940 b), hence LiF detectors consisting of natural lithium are sensitive to slow neutrons, their response to neutrons being enhanced by $^6$Li-enriched lithium or suppressed by using lithium consisting entirely of $^7$Li. Pairs of $^6$LiF/$^7$LiF detectors allow distinguishing between neutron/non-neutron components of radiation field.

For detection of neutrons of higher energies there is a need for moderators. Sixteen pieces of cylindrical moderators (see fig. 1a) have been produced from polyethylene PE-HD (PE-300) rods. Each moderator consists of moderator body (cylinder) and a plug (30 mm diameter) with detector box (6 mm height) mounted at the bottom of it (fig. 1b). Moderators were numbered (A1-A8, B1-B8). The diameter of moderator cylinders was 25 cm, while the cylinders height was 25 cm except for A7&A8 which were 21 cm.

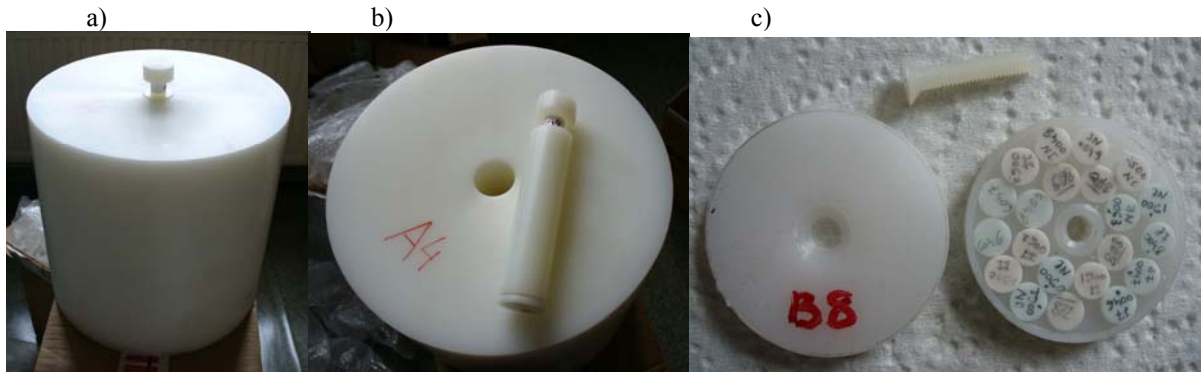

Fig. 1. a) The moderator, b) the plug with detectors' box mounted, c) the detectors' box filled in with TL detectors.

To prepare TL detectors for measurement the standard pre-irradiation annealing cycle was applied. Detectors' boxes have been filled in with detectors, three or four pieces of each of six types in each box (21-22 pcs in total, see fig. 1c), sixteen for A1-A8 and B1-B8 moderators, two for background (BG) measurement and one for transport dose (T) measurement. The boxes with detectors have been mounted at the bottom of plugs and inserted into moderator cylinders. Complete dosemeters have been sent to JET while some TLDs from each type were packed in polymethacrylate (PMMA) boxes and kept in low dose lead container/house at IFJ lab for calibration purposes and background evaluation.

**2.2 Measurements**

All dosemeters arrived at Culham early April 2012. A1-A8 and B1-B3 dosemeters inside the moderator cylinders were located by JET team in 11 positions in the Torus Hall and the two-week exposure took place in summer 2012 during the last phase of C30 experimental campaign. $1.21 \times 10^{18}$ neutrons were produced by JET plasmas during this period. Dosemeters A1, A7, A8 and B1 were situated close to the tokamak, A2, A3, A4, A5, A6 in the labyrinth located in the South West corner of Torus Hall, and B2 and B3 in the chimney located in the South East corner of Torus Hall. B4-B8 dosimeters were stored in J1D lab storage in PE boxes + plugs but not in cylinders. Finally, the two background and the transport dosimeters were located within their boxes in an office drawer for background measurement. An overall view of dosemeters positions is given in fig. 2. The dosemeters were removed in August 2012 and sent back to the IFJ in September 2012.

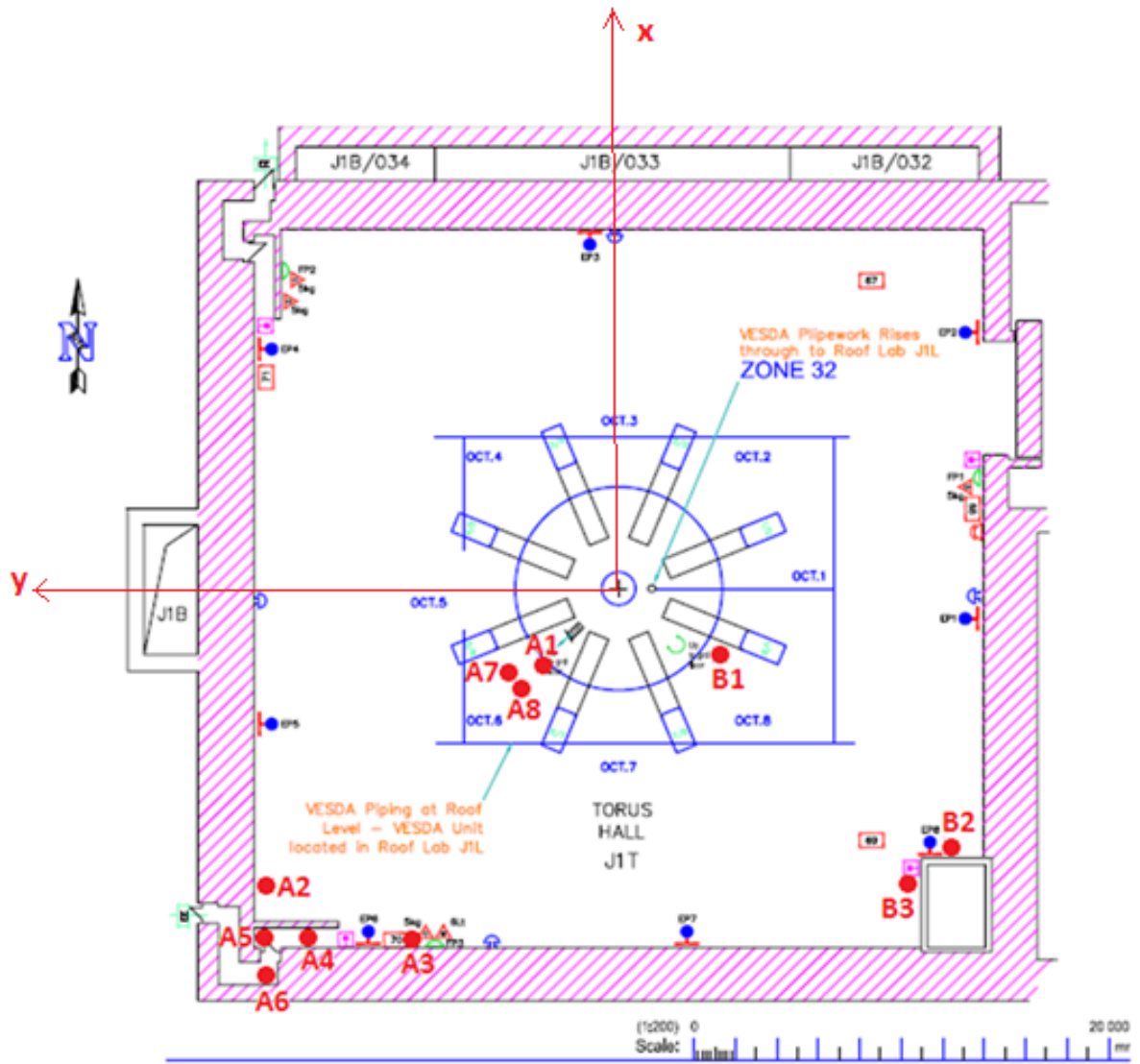

Fig. 2. Overview of TLDs location in the JET Torus Hall [12]. Note that detectors are at different heights.

**3. Results**

The TL signal measured for all detectors have been calibrated in terms of kerma in air with Cs-137 gamma rays. For gamma rays, it can be demonstrated that kerma in air is equivalent to dose in air under the charged particle equilibrium conditions provided by the PE boxes. Results of measurements taken at the different positions calculated as mean value of the signal of MCP and MTS detectors of different types are presented in fig. 3. Single measurement uncertainties were calculated ranging from a few percent of measured value up to about 11%.

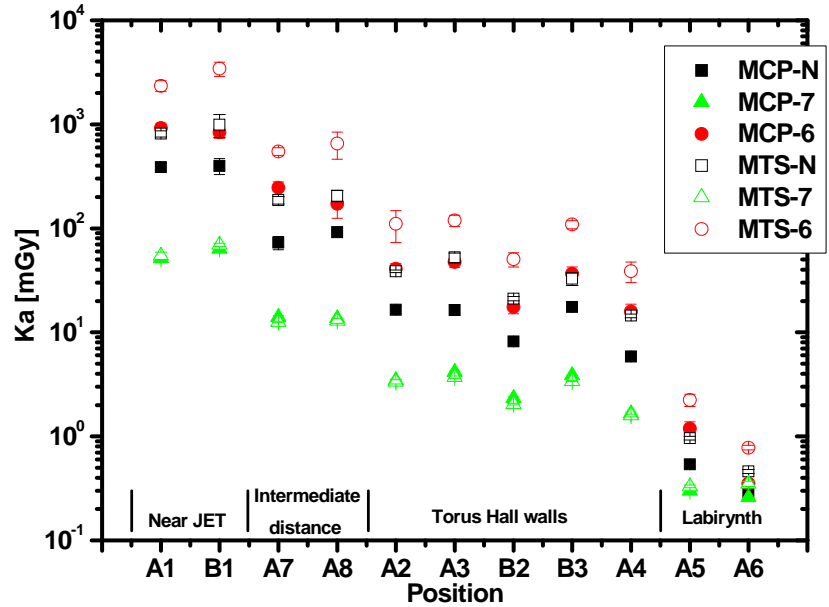

Fig. 3. Measurements results calculated as mean value of the same type detectors response calibrated in kerma in air with Cs-137 gamma rays.

As can be seen the measured absorbed doses decrease significantly with increasing distance from the tokamak (see fig. 2). Also, the doses resulted from response of A4, A5 and A6 dosemeters which were positioned inside the labyrinth are decreasing with decreasing distance to the labyrinth exit from the Torus Hall, as expected.

It is visible from the data presented above that the recorded dose evaluated from MTS-N detectors response is higher than from MCP-N (with a factor 2-3), while for Li-6 enriched MTS-6 and MCP-6 detectors a difference is even higher (up to 4 times). This is due to lower efficiency of detection of high-Linear Energy Transfer (LET) particles (among them thermal neutrons) by MCP detectors [12]. The self-shielding effect of neutrons by natural Li and $^6$Li enriched detectors must also be considered.

Unfortunately, only kerma in air values measured by $^7$Li enriched detectors showed relatively low dispersion (a few %) from the mean value calculated from all detectors of the same type. The results obtained from detectors containing higher amount of $^6$Li, i.e. $^6$Li enriched detectors, but also those produced from Li with natural abundance of isotopes, showed very high dispersion, in a few cases even close to 30%. That can be explained by observing that the presence of $^6$Li detectors have a shadowing effect on the surrounding detectors and therefore affect the measurements. So, for detectors which contain higher amount of $^6$Li, the maximum value measured at each position is the more reliable value. Due to this mean values for MCP-7/MTS-7 detectors and maximum measured values for MCP-6/MTS-6 and MCP-N/MTS-N detectors have been respectively used at each position for the next evaluation step.

As mentioned earlier, subtracting response of MTS-7 (or MCP-7) from the signal measured by MTS-6 or MTS-N detectors (or MCP-6 and MCP-N, respectively) it is possible to evaluate the TLDs response part due to the neutron component of the field. Evaluation of both components from MTS and MCP measurement for all positions used at JET is presented in fig. 4. MCP detectors type lower efficiency of neutron detection is well visible from these data.

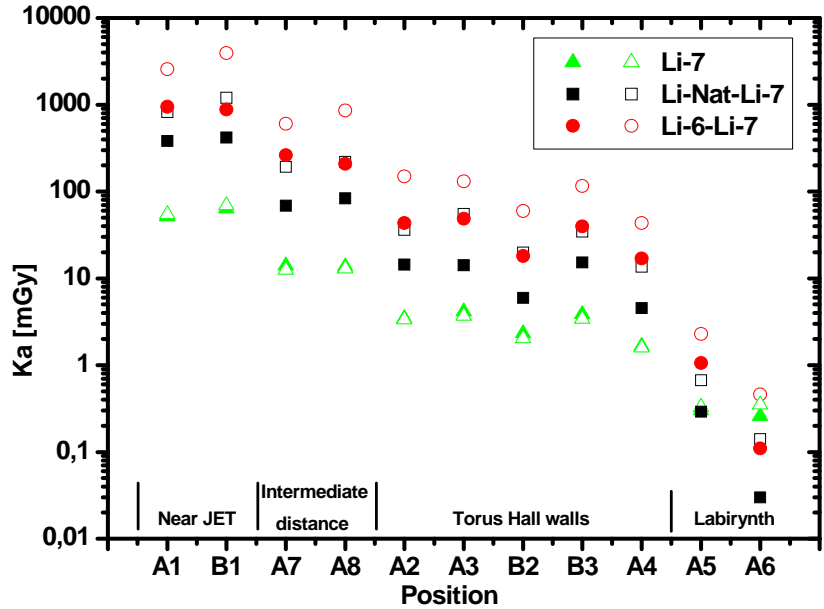

Fig. 4. Non-neutron and neutron component of the radiation field (for MCP detectors - filled symbols, for MTS - empty symbols).

It is worth noting that, while calibrating TLDs in terms of gamma kerma in air is straightforward using calibration gamma sources, it is not so for TL neutron signal and therefore, usually, the response of TLDs due to the neutron components is related to the neutron fluence in a well defined neutron energy spectrum, which is measurable. In particular, the TLDs response due to neutron component of the radiation field can be related to the local neutron fluence taking into account LiF detectors' calibration at the PTB Thermal Neutron Reference Field at GeNF [13] performed in 2006 by Burgkhardt et al. [14]. This calibration data can be regarded as results for pure thermal neutron field.

Thanks to this calibration we can evaluate the neutron fluence for detectors of each type at every measurement location. In fact, in the case of JET measurements, although the neutron spectrum is not thermal in any position in the Torus Hall, the large PE moderators ensures that the enclosed TLDs "see" a pure thermal neutron field, and hence the Burgkhardt calibration factors can be applied. The resulting values of this evaluation are shown in fig. 5.

Some dosimeters (B4-B8) were stored in J1D lab storage, while BG and T dosimeters were located in an office drawer for background measurement. The difference between MCP-N and MCP-7 detectors signal in $K_a$ for B4-B8 is about 0.02 mGy, while for BG&T about 0.01 mGy. Using calibration made at the PTB thermal neutron source, neutron fluence can be estimated roughly as $5 \times 10^5$ n/cm$^2$ neutron background for B4-B8 in J1D lab while about half of it for BG&T in the office, due to natural background. It has to be bear in mind, however, that B4-B8 were kept screwed to plugs while BG&T were naked, neither were contained in PE cylinders.

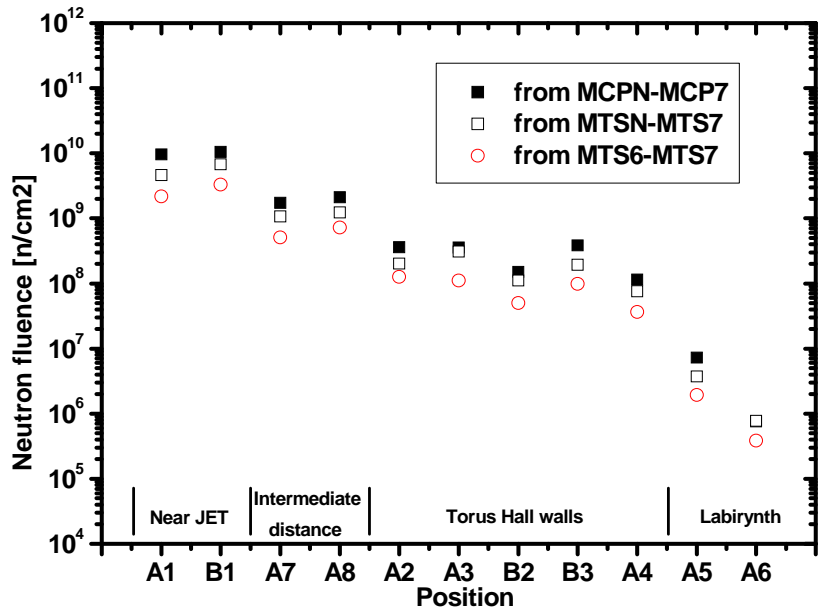

Fig. 5. Neutron fluence at different positions evaluated from response of detectors with different Li-6 content.

## 4. Simulations

The objective of numerical simulations was to calculate the neutron fluence at detectors and compare it with the measured fluence by the TLDs, in particular by MTS-N, MTS-6 and MCP-N, which are calibrated in terms of neutron fluence. Calculations were performed using Monte Carlo code MCNPX 2.5 [15]. The existing model of JET torus and of the Torus Hall [16] was used to calculated the neutron fluence at detectors located close to the machine (A1, B1, A7-A8), and on the SE corner of the hall (B2-B3). For the detectors located close to the labyrinth and inside it (A2-A6), a stage-by-stage simulation approach was employed. The model of the JET torus was used to produce a surface neutron source. The Surface Source Write (SSW) file registered neutrons on a quarter sphere with centre at the South West Torus Hall corner (1.0 m above the floor surface) and radius of 5.0 m. Contribution of neutrons leaking from the torus and scattered in the wall materials was taken into account. The SSW file was used as Surface Source Read (SSR) input file for the calculations performed for detectors in the labyrinth area. Neutron fluence and ambient dose equivalent were calculated along the total length of the maze. Cross-section data were obtained from FENDL-2.1 and ENDF-VI-8.

The experimental set up (fig.1) was also modelled by MCNPX. Neutron self-shielding and interferences between crystals were estimated for different types of TLDs used for all test positions. Neutron self-shielding depends on TL detector material, dimensions, and geometry and the neutron energy spectrum. Self-shielding of each TLD is already taken into account in the calibration at the given calibration conditions. However, as the TLDs were calibrated individually, in the present case it is important to take into account the shielding effect of the simultaneous presence of about 20 TLDs in the same box. The perturbation factor of the neutron field due to the interfering presence of the TLDs and for the specific configuration examined was found to be of about 0.9. Finally, the experimental results were compared with calculations and the C/E values were obtained.

C/E values for MCP-N TLDs are shown in fig. 6. As a general result, calculations underestimate the measurements in positions close to the machine and overestimate them in positions far from the machine. The only exception is observed for the A3 detectors for which calculation underestimates the neutron fluence (note that A3 has the same measured fluence as A2, but much lower calculated fluence, the two detectors are located at different heights).

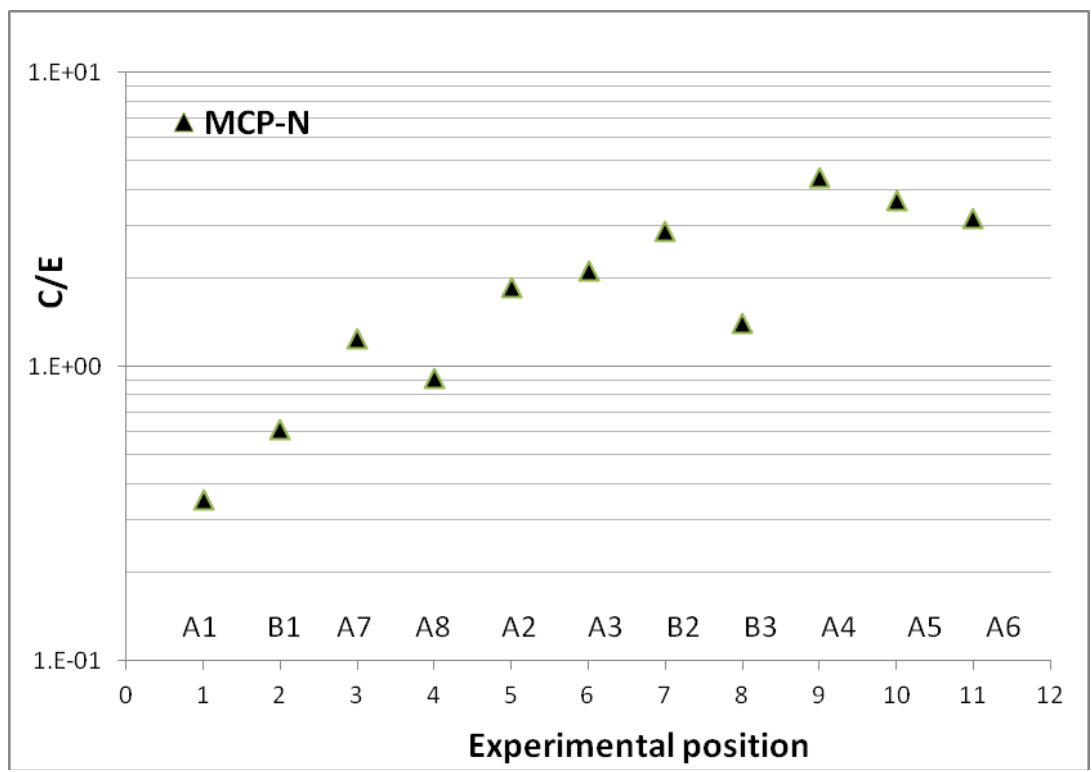

Fig. 6. C/E values for the neutron fluence obtained using MCP-N detectors as a function of experimental position.

The observed discrepancies in C/E ratios can be attributed to several factors:

(a) Approximations in the calculation of the neutron fluence at detector positions: MCNP model of JET has been developed to calculate the neutron flux and fluence inside the machine and outside to the magnetic limbs. JET machine is described to sufficient details in the MCNP model but the large diagnostic systems, heating systems and various equipments surrounding the machine are not described in detail in the MCNP model. These have probably a shielding effect on positions far from the machine.

(b) The TLDs were calibrated in a thermal neutron spectrum. The neutron spectrum is not fully thermalized by the use of polyethylene cylinders in positions close to the tokamak. In these positions, TLDs experience a significant fraction of 2.45 MeV neutrons and the TLDs calibration factors are therefore not correct in these positions.

(c) Shielding and shadowing effects between TLDs could not be accurately evaluated because the orientation of cylinders with respect to the neutron source was not recorded, and because some of the TLDs were found dislocated from their original locations, and overlapped, after exposure. According to the analysis performed, however, the observed discrepancies between calculated and experimental neutron fluence values cannot be attributed to the shielding and shadowing effect.

(d) The largest discrepancies were found for Li-6 enriched TLDs, which have the largest self-shielding effect.

Although the errors in the measurements and the statistical uncertainties in the calculations are small, the total uncertainty in the C/E comparison cannot be easily quantified because of the circumstances discussed above, but is expected to be large.

**5. Summary and conclusions**

The neutron fluence during JET plasma discharges has been measured in various positions in the Torus Hall, close to the JET machine and in the labyrinth. The fluence in these positions varied over more than four orders of magnitude. The measurement was possible thanks to the use of different types of very sensitive TLDs. As these TLDs contain different amount of $^6$Li and $^7$Li, the different contributions of neutrons and of gamma rays to the total dose could be separated. Moreover, as the TLDs are also calibrated in terms of neutron fluence, the local neutron fluence could be obtained from the neutron dose measurements. Measurements of the gamma dose and of the neutron fluence were obtained for all positions over a range of about five orders of magnitude variation.

These measurements are very promising and have shown that a more extensive mapping of the neutron fluence further in the labyrinth and in the chimney down to the Torus Hall basement can be obtained during a JET campaign with substantial neutron production.

As a general result, and compared with the measured values, calculations underestimate the measurements in positions close to the machine and overestimate them in positions far from the machine. The possible reasons for the observed discrepancies were discussed. We note that this is almost the first attempt to characterize the neutron flux in the JET Torus Hall. A previous such experiment, carried out during the DTE1 campaign using activation foils, had obtained similar results [17]. Despite the large discrepancies in calculated C/E ratios, the first results of this work can be considered as satisfactory if one takes into account the complexity of the actual JET tokamak, experimental hall and shielding configurations as well as TLD calibration and signal interpretation procedures employed.

The results confirm that the TLD technology can be usefully applied to measurements of neutron streaming through JET Torus Hall ducts. New detector positions, further in the labyrinth and ducts, will be investigated in the next measurement campaigns. The number, type and positioning of detectors inside the moderators will be improved to reduce the shadowing effect observed for detectors containing $^6$Li.

The results of this work assist operational radiation protection activities in the JET facility. Moreover, the comparison of the results of the present computations against experimental measurements enables validation of the numerical tools used for ITER design. Therefore, the experiences acquired during JET operation will provide a firm base for implementation of this technology in area radiation monitoring of ITER aiming to minimize personnel dose in accordance to the ALARA principle.

**Acknowledgments**

This work was supported by EURATOM and carried out within the framework of the European Fusion Development Agreement (EFDA), under JET Fusion Technology task JW12-FT-5.45.